\documentstyle[epsf]{elsart}
\begin{document}

\def\spose#1{\hbox to 0pt{#1\hss}}
\def\ltapprox{\mathrel{\spose{\lower 3pt\hbox{$\mathchar"218$}}
 \raise 2.0pt\hbox{$\mathchar"13C$}}}
\def\gtapprox{\mathrel{\spose{\lower 3pt\hbox{$\mathchar"218$}}
 \raise 2.0pt\hbox{$\mathchar"13E$}}}
\def\inapprox{\mathrel{\spose{\lower 3pt\hbox{$\mathchar"218$}}
 \raise 2.0pt\hbox{$\mathchar"232$}}}

\begin{frontmatter}

\begin{flushright}
{\normalsize FSU-SCRI-99-31}\\
{\normalsize hep-lat/9905008}\\
\end{flushright}

\title{Thermodynamics with Dynamical Clover Fermions}
\author{
Robert G. Edwards and Urs M. Heller}
\address{
SCRI, The Florida State University, 
Tallahassee, FL 32306-4130, USA}

\begin{abstract}
We investigate the finite temperature behavior of nonperturbatively
improved clover fermions on lattices with temporal extent $N_t=4$ and $6$.
Unfortunately in the gauge coupling range, where the clover coefficient
has been determined nonperturbatively, the finite temperature
crossover/transition occurs at heavy pseudoscalar masses and large
pseudoscalar to vector meson mass ratios. However, on an $N_t=6$ lattice
the thermal crossover for the improved fermions is much smoother than
for unimproved Wilson fermions and no strange metastable behavior is
observed.
\end{abstract}
\end{frontmatter}


Simulations with Wilson fermions suffer ${\cal O}(a)$ scaling violations
due to the dimension five operator that Wilson introduced to give the
unwanted fermion doublers masses of order the cutoff, $1/a$. These
scaling violations are much larger than those in the glue sector, which
are ${\cal O}(a^2)$, and they can be numerically quite large,
necessitating use of small lattice spacings at large simulation cost to
get results that can be reliably extrapolated to the continuum limit.

The ${\cal O}(a)$ scaling violations can be reduced to ${\cal O}(a^2)$ by
introducing another dimension five operator into the fermion action,
the so called clover term,
\begin{equation}
S_{\rm sw} = c_{\rm sw} \kappa \sum_x \bar \psi(x) i \sigma_{\mu \nu}
{\cal F}_{\mu \nu}(x) \psi(x) ~,
\label{eq:clov}
\end{equation}
where ${\cal F}_{\mu \nu}(x)$ is a lattice transcription of the field
strength tensor $F_{\mu \nu}(x)$, usually taken from four open plaquettes
looking like a clover leaf, as proposed by Sheikholeslami and
Wohlert~\cite{SW}. For the reduction of scaling violations to work
the clover coefficient, $c_{\rm sw}$, needs to be determined
nonperturbatively as a function of the gauge coupling. The ALPHA
collaboration developed a method to do so within the Schr\"odinger
functional framework~\cite{ALPHA}. For quenched QCD the nonperturbative
clover coefficient is now known for gauge coupling $6/g^2 \ge 5.7$,
corresponding to lattice spacings $a \ltapprox 0.17$ fm~\cite{ALPHA,EHK}.
A substantial reduction of scaling violations in this region from
${\cal O}(a)$ to ${\cal O}(a^2)$ has been verified nicely in \cite{EHK}.

The clover coefficient has recently also been determined by the ALPHA
collaboration for full QCD with two flavors of dynamical fermions for
gauge coupling $\beta = 6/g^2 \ge 5.2$, corresponding roughly to
lattice spacings $a \ltapprox 0.14$ fm~\cite{JS}. To be precise, the
clover coefficient was determined for $\beta \ge 5.4$ and fitted to
a ratio of polynomials in $g^2$. At $\beta = 5.2$ a numerical consistency
check with the value extrapolated with this function of $g^2$ was
performed (see~\cite{JS} for details). Preliminary first results of
hadron spectroscopy and the heavy quark potential in dynamical
simulations with this nonperturbatively determined clover coefficient
have appeared in~\cite{UKQCD}.

Arguably the largest lattice artifacts in simulations with dynamical
Wilson fermions have been observed in simulations probing the finite
temperature behavior in the vicinity of the deconfinement/chiral
symmetry restoration transition or (at finite quark mass) crossover.
The most likely scenario for the behavior of two-flavor QCD at finite
temperature is a rapid crossover for finite quark mass, turning into
a second order chiral symmetry restoring phase transition in the
massless limit~\cite{PW}. However, simulations with dynamical Wilson
fermions showed strange, unexpected behavior, {\it e.g.} first order phase
transition like signals at intermediate quark masses, that softened again
at smaller quark masses~\cite{MILC_thWi,MILC_thWi2}.

It has been argued~\cite{Iwasaki_97} that this strange and unexpected
behavior is due to effects of the Wilson pure gauge action in its
so called ``crossover region'', where the plaquette varies sharply
with changes in the gauge coupling, which feeds back to the fermions,
rather than being due to artifacts in the fermion action. Indeed,
their simulations did not show any evidence for first order like
signals. Similarly, a study using both an improved gauge and a
clover improved fermion action, with so-called tadpole improved
coefficients, found a smoother behavior in the thermal crossover
region than the simulations with unimproved Wilson action for both
gauge and fermion sectors~\cite{MILC_clov}. From the point of view
of the Symanzik improvement program both these simulations still
have ${\cal O}(g^n a)$ errors of unknown magnitude. Furthermore,
it is not clear, in the study where both gauge and fermion action
are improved, which improvement is more important in smoothening
out the thermal crossover behavior.

In this letter, we study the effect of the nonperturbative improvement
of the Wilson fermion action on the behavior at the finite temperature
crossover. We are interested in simulations with small lattice extent
in the temporal direction, {\it i.e.} at large lattice spacing, where
the simulations are relatively cheap.

\begin{figure}
\epsfxsize=3.5in
\centerline{\epsfbox[125 80 525 490]{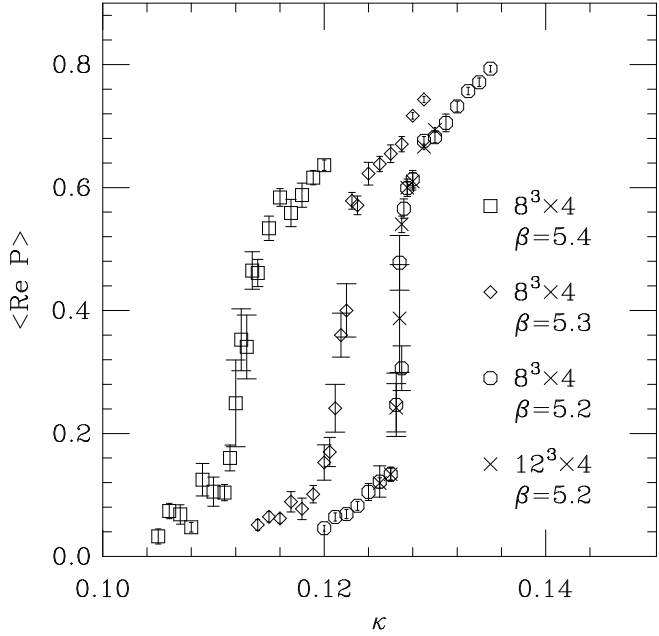}}
\caption{$\langle {\mathrm Re} P \rangle$ for the simulations with
$N_t=4$ and volume $8^3$ at $\beta=5.4$ (squares) $5.3$ (diamonds)
and $5.2$ (octagons) and with volume $12^3$ at $\beta=5.2$ (crosses).}
\label{fig:ReP_nt4}
\end{figure}
\begin{figure}
\epsfxsize=3.5in
\centerline{\epsfbox[125 80 525 490]{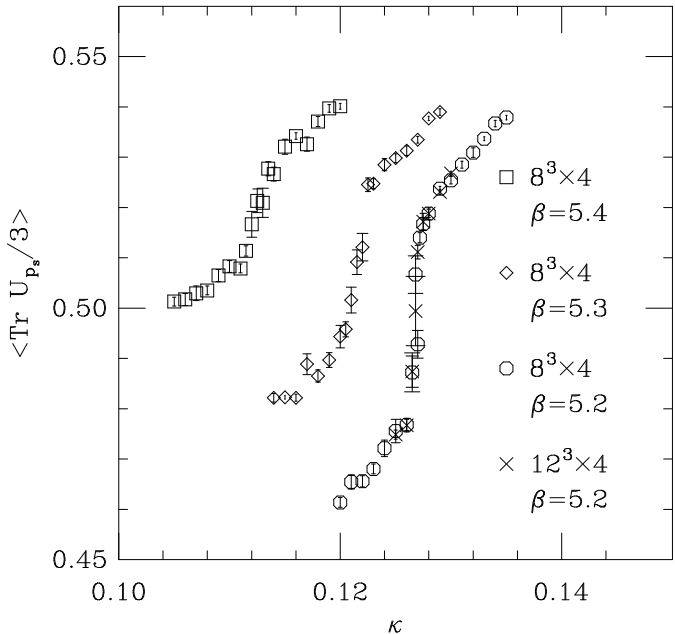}}
\caption{Space-like plaquettes, $\langle {\mathrm Tr} U_{p_s} / 3 \rangle$,
for the same lattices and with the same symbols as
Fig.~\protect\ref{fig:ReP_nt4}.}
\label{fig:plaq_nt4}
\end{figure}

The largest coupling for which the nonperturbative value of $c_{sw}$ is
known is $\beta = 6/g^2 = 5.2$. We performed simulations on $8^3 \times 4$
lattices for $\beta=5.4$, $5.3$ and $5.2$ and various values of $\kappa$
in the thermal transition/crossover region. The values for $c_{sw}$ used,
obtained from Ref.~\cite{JS} are listed in Table~\ref{tab:csw}. Measurements
were taken, after thermalization, over 500 trajectories away from the
crossover region, and over up to 4000 trajectories in the middle of the
crossover region. We show in Figure~\ref{fig:ReP_nt4} the real part of the
Polyakov line expectation value and in Figure~\ref{fig:plaq_nt4} the average
space-like plaquette. A clear crossover is seen for all three gauge
couplings, becoming sharper as the coupling is increased (as $\beta$ is
decreased). For the largest coupling, $\beta=5.2$, we also simulated on a
larger spatial volume, $12^3 \times 4$. As can be seen from the figures,
there is no evidence for finite volume effects.

\begin{table}
\caption{The clover coefficients, $c_{sw}$, used in the simulations and
estimates of the $\kappa_c$'s. They were obtained from Ref.~\protect\cite{JS}.}
\vskip 0.3cm
\label{tab:csw}
\begin{center}
\begin{tabular}{|l|l|l|} \hline
 $\beta$ & $c_{sw}$ & $\kappa_c$ \\ \hline
 5.4 & 1.82277 & 0.1370 \\
 5.3 & 1.90952 & 0.1370 \\
 5.2 & 2.02    & 0.1370 \\ \hline
\end{tabular}
\end{center}
\end{table}

The thermal crossover appears quite rapid, reminiscent of unimproved
Wilson simulations~\cite{MILC_thWi}, as compared to thermodynamics with a
tadpole improved clover fermion and Symanzik improved gauge
action~\cite{MILC_clov}. But, of course, comparisons should not be made
in terms of bare parameters, such as $\kappa$. We have therefore computed
hadron masses and the heavy quark potential at the thermal crossover points.
For $\beta=5.2$ hadron masses were also computed on both sides of the
thermal crossover. Most of these measurements were done on $8^3 \times 16$
lattices. In one case, for $\beta=5.2$ and $\kappa=0.1340$, where the meson
masses are lightest, we also have preliminary results from a simulation on
a larger $16^3 \times 32$ lattice. There, we do see finite size effects on
the masses of about 10\%, but much less on the mass ratio. We suspect that
the masses for the next lightest quark mass, at $\beta=5.2$, $\kappa=0.1330$
are also affected by some small finite size effects, while for the other
cases the finite size effects are expected to be smaller than the statistical
errors. All the results are collected in Table~\ref{tab:spectro}.

\begin{table}
\caption{Masses of pseudoscalar and vector mesons and their ratio, and
values for the string tension from $L^3 \times (2L)$ lattices. Points
marked by an asterisk lie on the $N_t=4$ crossover, and the point marked
by a plus on the $N_t=6$ crossover.}
\vskip 0.3cm
\label{tab:spectro}
\begin{center}
\begin{tabular}{|r|c|r|l|l|l|l|} \hline
 $\beta$ & $\kappa$ & $L$ & $am_{PS}$ & $am_V$ & $m_{PS}/m_V$ &
 $a\sqrt{\sigma}$ \\
\hline
 $^\ast$5.4 & 0.1125 &  8 & 2.167(3) & 2.237(4)  & 0.968(1)  & 0.479(18) \\
 $^\ast$5.3 & 0.1215 &  8 & 1.833(3) & 1.948(4)  & 0.941(1)  & 0.530(7)  \\
 5.2        & 0.1260 &  8 & 1.656(3) & 1.814(5)  & 0.913(2)  &           \\
 $^\ast$5.2 & 0.1270 &  8 & 1.594(2) & 1.757(3)  & 0.907(1)  & 0.559(10) \\
 5.2        & 0.1280 &  8 & 1.506(3) & 1.672(7)  & 0.901(3)  &           \\
\hline
 5.2        & 0.1320 &  8 & 1.026(5) & 1.185(27) & 0.866(18) &           \\
 $^+$5.2    & 0.1330 &  8 & 0.793(5) & 0.934(7)  & 0.849(5)  & 0.304(7)  \\
 5.2        & 0.1340 &  8 & 0.691(4) & 0.877(4)  & 0.789(3)  & 0.283(12) \\
 5.2        & 0.1340 & 16 & 0.631(1) & 0.787(3)  & 0.802(3)  & 0.290(8)  \\
\hline
\end{tabular}
\end{center}
\end{table}

\begin{figure}
\epsfxsize=3.5in
\centerline{\epsfbox[125 80 525 490]{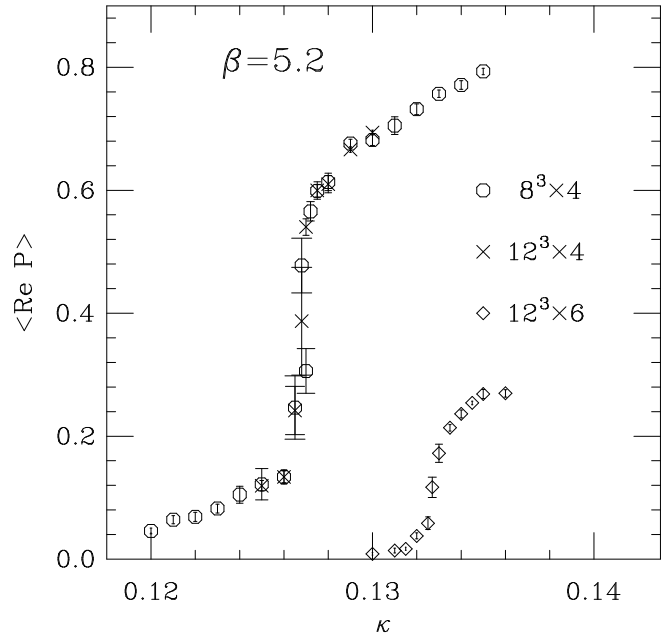}}
\caption{$\langle {\mathrm Re} P \rangle$ for the simulations at $\beta=5.2$
for lattice size $8^3 \times 4$ (octagons), $12^3 \times 4$ (crosses)
and $12^3 \times 6$ (diamonds).}
\label{fig:ReP_5p2}
\end{figure}

\begin{figure}
\epsfxsize=3.5in
\centerline{\epsfbox[125 80 525 490]{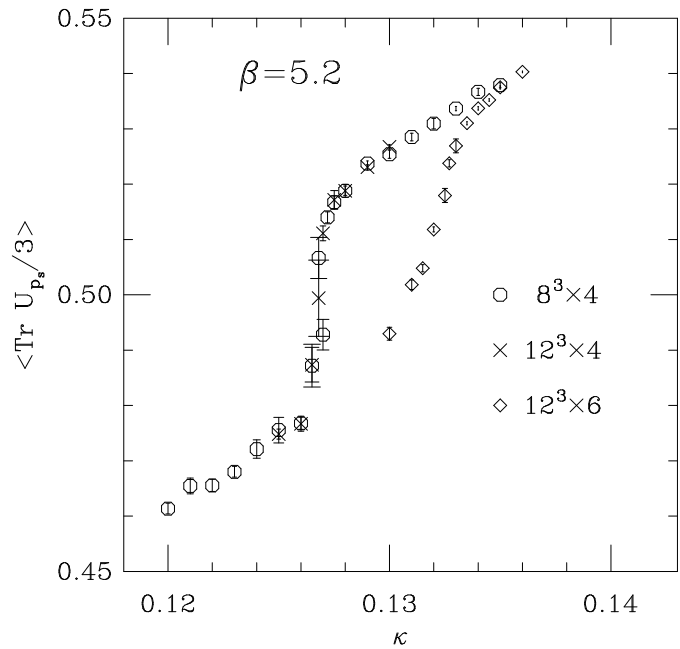}}
\caption{Space-like plaquettes, $\langle {\mathrm Tr} U_{p_s} / 3 \rangle$,
for the same lattices and with the same symbols as
Fig.~\protect\ref{fig:ReP_5p2}.}
\label{fig:plaq_5p2}
\end{figure}

Even at the strongest coupling, for which the nonperturbative clover
coefficient is known, the thermal crossover for $N_t=4$ lattices occurs
at heavy pseudoscalar meson mass and large pseudoscalar to vector meson
mass ratio. We therefore also considered thermodynamics on an $N_t=6$
lattice at $\beta=5.2$. $\langle {\mathrm Re} P \rangle$ and $\langle
{\mathrm Tr} U_{p_s} / 3 \rangle$ are shown in Figures~\ref{fig:ReP_5p2}
and \ref{fig:plaq_5p2}, where they are compared to the results from
$N_t=4$. The crossover seems somewhat smoother for $N_t=6$, and it occurs
at lighter meson masses. However, at $0.85$ the pseudoscalar to vector
meson mass ratio is still rather large.

\begin{figure}
\epsfxsize=3.5in
\centerline{\epsfbox[125 80 525 490]{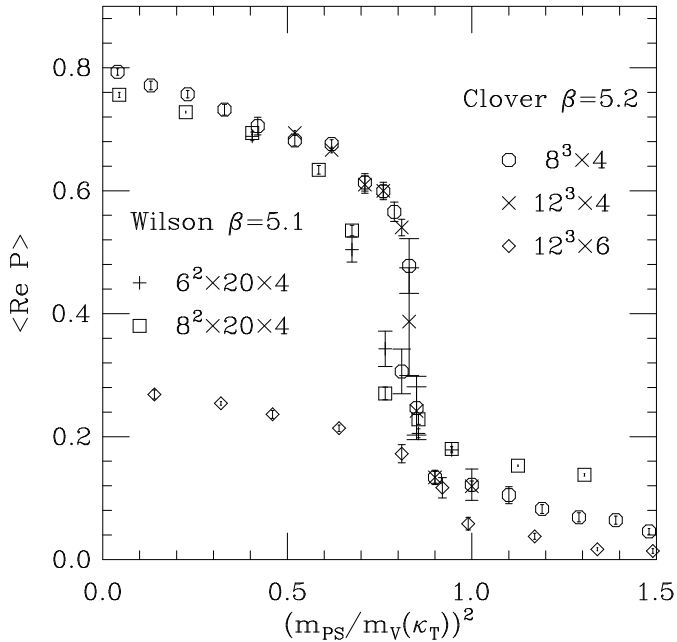}}
\caption{Comparison of $\langle {\mathrm Re} P \rangle$ for the simulations
with clover improved and non-improved Wilson fermions at similar
$m_{PS}/m_V$ ratios at the thermal crossover as function of
$(m_{PS}/m_V(\kappa_T))^2$, giving a comparable measure of the
quark masses, for $N_t=4$. Also shown are the clover results for
$N_t=6$.}
\label{fig:ReP_cl_wl}
\end{figure}

Comparing Figure~\ref{fig:ReP_nt4} with similar plots for unimproved
Wilson fermions such as Fig.~5 of \cite{MILC_thWi} the crossover for
the improved clover fermions appears to be even more pronounced than
for the unimproved Wilson fermions. Comparing as function of the bare
hopping parameter $\kappa$, though, can be misleading and we therefore
follow the strategy of Ref.~\cite{MILC_clov} and make a comparison as
function of the quark mass, or equivalently as function of $m_{PS}^2$. 
{}From Ref.~\cite{HEMCGC_th} we can see that the crossover for Wilson
fermions at $\beta=5.12$ and $\kappa=0.1700$ occurs at a pseudoscalar
to vector meson mass ratio $m_{PS}/m_V=0.899(4)$. It appears that the
unimproved Wilson fermion data at $\beta=5.1$ of \cite{MILC_thWi}
are the set that most closely matches ours at $\beta=5.2$. In addition,
from Ref.~\cite{beta_fn} we have some mass measurements in the thermal
crossover region. While the $m_{PS}/m_V$ ratios at the thermal crossover
are comparable, the masses in lattice units are quite different. We
therefore decided to plot $\langle {\mathrm Re} P \rangle$ as function
of $(m_{PS} / m_V(\kappa_T))^2$, where $m_V(\kappa_T)$ is the vector
meson mass at the thermal crossover point $\kappa_T$. The comparison is
shown in Fig.~\ref{fig:ReP_cl_wl}. It appears that the crossover for
the improved fermions with $N_t=4$ is somewhat sharper than for the
unimproved fermions.

The crossover for the improved fermions on the $N_t=6$ lattice, on the
other hand, is much smoother. The pseudoscalar to vector meson mass ratio
at the crossover corresponds to that for unimproved Wilson fermion
at $\beta=5.22$, $\kappa=0.17$~\cite{MILC_thWi2}. This is the region
where first order like metastable states were observed in \cite{MILC_thWi2}.
So here the improvement seems to help.

In conclusion, we have carried out finite temperature simulations with
nonperturbatively improved Wilson fermion in the region of largest coupling,
and hence largest lattice spacing, for which the nonperturbative value
of the clover coefficient $c_{sw}$ is known. The thermal crossover on
$N_t=4$ lattices occurs at very heavy pseudoscalar meson masses and large
pseudoscalar to vector meson mass ratios. The crossover appears somewhat
sharper than for unimproved Wilson fermions at comparable $m_{PS}/m_V$
ratios. However, for the improved fermions the masses (in lattice units)
are considerably larger, and the thermal crossover could still be
significantly influenced by the deconfinement transition in the pure
gauge theory. While the thermal crossover on an $N_t=6$ lattice at the
strongest accessible coupling still occurs at a large $m_{PS}/m_V$
ratio, it has become much smoother, in particular compared to unimproved
Wilson fermions, which show strange first order like behavior for
comparable $m_{PS}/m_V$ ratios. To get the thermal crossover to occur
for smaller $m_{PS}/m_V$ ratio, one needs to use lattices with larger
temporal extent $N_t$ or try to push the nonperturbative determination of
$c_{sw}$ to stronger couplings.

\ack{
This research was supported by DOE contracts
DE-FG05-85ER250000 and DE-FG05-96ER40979.  Computations were performed
on the CM-2 and the workstation cluster at SCRI.
}


\begin{thebibliography}{9}

\bibitem{SW} B. Sheikholeslami and R. Wohlert,
  {\em Nucl. Phys. \/}{\bf B259} (1985) 572.
\bibitem{ALPHA} M. L\"uscher, S. Sint, R. Sommer, P. Weisz, and U. Wolff,
  {\em Nucl. Phys. \/}{\bf B478} (1996) 365,
  {\em Nucl. Phys. \/}{\bf B491} (1997) 323, 344.
\bibitem{EHK} R.G. Edwards, U.M. Heller and T.R. Klassen,
  {\em Phys. Rev. Lett. \/}{\bf 80} (1997) 3448.
\bibitem{JS} K. Jansen and R. Sommer,
  {\em Nucl. Phys. \/}{\bf B530} (1998) 185.
\bibitem{UKQCD} M. Talevi (The UKQCD Collaboration), hep-lat/9809182;
  C.R. Allton {\em et al.} (The UKQCD Collaboration), hep-lat/9808016.
\bibitem{PW} R. Pisarski and F. Wilczek, {\em Phys. Rev. \/}{\bf D29}
  (1984) 339.
\bibitem{MILC_thWi} C. Bernard {\em et al.} (The MILC collaboration),
  {\em Phys. Rev. \/}{\bf D49} (1994) 3574.
\bibitem{MILC_thWi2} C. Bernard {\em et al.} (The MILC collaboration),
  {\em Phys. Rev. \/}{\bf D46} (1992) 4741;
  T. Blum {\em et al.} (The MILC collaboration), {\em Phys. Rev. \/}{\bf D50}
  (1994) 3377.
\bibitem{Iwasaki_97} Y. Iwasaki, K. Kanaya, S. Kaya and T. Yoshi\'e,
  {\em Phys. Rev. Lett. \/}{\bf 78} (1997) 179.
\bibitem{MILC_clov} C. Bernard {\em et al.} (The MILC collaboration),
  {\em Phys. Rev. \/}{\bf D56} (1997) 5584.
\bibitem{HEMCGC_th} K.M. Bitar {\em et al.} (The HEMCGC Collaboration),
  {\em Phys. Rev. \/}{\bf D43} (1991) 2396.
\bibitem{beta_fn} K.M. Bitar, R.G. Edwards, U.M. Heller and A.D. Kennedy,
  {\em Phys. Rev. \/}{\bf D54} (1996) 3546.

\end{thebibliography}
\end{document}